# 6

# The Six Ways to Build Trust and Reduce Privacy Concern in a Central Bank Digital Currency (CBDC)

**Alex Zarifis and Xusen Cheng**

## 6.1 Introduction

Central Bank Digital Currencies (CBDCs) have been implemented by only a handful of countries, but they are being explored by many more (Kosse & Mattei, 2022). CBDCs are digital currencies issued and backed by a central bank. Consumer trust can encourage or discoura ge the adoption of this currency, which is also a payment system and a technology. This research attempts to understand consumer trust in CBDCs so that the development and adoption stages are more effective and satisfying for all the stakeholders.

_______________________

A. Zarifis (✉)
School of Business, University of Nicosia, Nicosia, Cyprus

Cambridge Centre for Alternative Finance, University of Cambridge, Cambridge, UK
e-mail: zarifis.a@unic.ac.cy

X. Cheng
School of Information, Renmin University of China, Beijing, China
e-mail: xusen.cheng@ruc.edu.cn

**115**

The digital transformation sweeping through many sectors of the economy and facets of people's lives is also disrupting money and payment systems. Governments want a currency and payment solution that is more efficient and intelligent, and consumers want to make transactions faster and cheaper (Abramova et al., 2021). Consumers have shown an interest in the functionality provided by cryptocurrencies, but the recurring problems with cryptoassets, such as the FTX collapse, suggest that a more centralized and regulated cryptocurrency solution should be provided by governments to offer similar functionality with less risk. If this is not achieved however, history shows that when consumers do not want a currency or payment system, they chose alternatives such as other countries' currencies, cryptoassets, Decentralized Finance (DeFi), or the grey economy (Grassi et al., 2022). Therefore, as with many solutions involving technology, it is more effective to design something appealing to the consumer than push a flawed solution onto them, either with promotions, regulations, or laws. It may seem like a simple choice to implement CBDCs, but however, it is far from simple. The cautious, slow progress of some central banks through many pilot implementations is probably warranted (Xu, 2022). Unlike some technologies like the metaverse that can be repeatedly pitched to the consumer year after year until they are successfully adopted, a failed implementation of a CBDC would have far-reaching consequences. This cautious approach, however, must not take longer than necessary, and the optimum solutions for CBDCs, including trust, must be reached. Recent events showed that governments around the world want to make a variety of interventions as fast as possible, such as giving citizens of a whole country some additional money to deal with high inflation (Walker & Sexton, 2022). Many consumers lives seem to be going from one disruption to the other recently (Gielens, 2022), with the word 'permacrisis' being selected as the word of the year by Collins Dictionary (Shariatmadari, 2022). The uncertainty many consumers face, and will probably face for some time to come, means having their trust should not be taken as a given. Research into CBDCs recognizes that more needs to be done to understand this phenomenon and, in particular, the role the technology plays from the consumer's perspective (Bhaskar et al., 2022). Consumer trust is an important factor in technology adoption (Lankton et al., 2015). Therefore, the two related research questions are:

*What are the methods to build consumer trust in a CBDC?*
*Does trust in a CBDC encourage its adoption by a consumer?*

There are a variety of CBDC solutions. The primary distinctions are between wholesale and retail CBDCs, and one or two-tier CBDCs (Auer & Böhme, 2020). For this research we are considering the one-tier CBDC solution. This is for two reasons. Firstly because of its popularity within early movers like the Chinese CBDC, formally known as the electronic Chinese Yuan or the e-CNY (Xu, 2022). The second reason is because it is a simpler implementation, with fewer factors influencing the consumer's perspective, and therefore easier to create a representative model and validate it.

This research found support for a model with six factors influencing trust in a CBDC and found that trust in the CBDC did indeed encourage its adoption. The six factors that influence trust in a CBDC identified in this research are the following: (1) Trust in the government and the central bank issuing a CBDC, (2) expressed guarantees for the user of a CBDC, (3) positive reputation of existing CBDCs implemented in other countries, (4) automation and reduced human involvement achieved by a CBDC technology, (5) trust-building functionality of a CBDC, and (6) privacy features of the CBDC wallet app and back-end processes such as anonymity.

The following literature review of CBDCs and trust gives a sufficient foundation for the research model, that is presented in the third section. This is followed by the methodology section, where the quantitative approach using Structural Equation Modelling (SEM) is outlined. After that comes the analysis and results, the discussion of the findings, and finally the conclusion, which includes suggestions for future research.

## 6.2  Theoretical Foundation

This review first explores the various models and implementations of CBDCs before moving on to trust in the related areas of currencies, payments, and technology. While the literature of trust in CBDCs does not sufficiently cover this topic, the extensive literature of trust in the constituent parts of a CBDC (currency, payment system, and technology) gives this research a sufficient theoretical foundation to build on.

## Central Bank Digital Currencies (CBDCs)

CBDCs are different to other cryptocurrencies like Bitcoin as they are issued by a country's central bank and cannot be mined. Therefore, they are very centralized and not decentralized at all. The use of cash is declining over time in most countries, and the idea to offer a purely digital currency and payment system is not new (Tee & Ong, 2016). As is often the case with new technologies, the timing can make the difference between successful adoption by consumers or failure. The confluence of cryptoasset popularity, DeFi popularity, the digital transformation in many sectors of the economy and several other factors have convinced many countries that the right time is now (Luu et al., 2022; Lee et al., 2021).

There are several variations of CBDCs. The most important variations are between (a) retail and wholesale and (b) one and two tiers (Auer & Böhme, 2020): (a) A CBDC can either be retail or wholesale. A retail CBDC is used by individuals for their savings and to make purchases. A wholesale CBDC is primarily used by financial institutions for their reserves and large payments. (b) The operations of a CBDC can either be over one or two tiers. If it operates over one tier, a central bank provides an electronic wallet to the individual to use it. The Chinese e-CNY operates in this way (Xu, 2022). A CBDC that operates with two tiers essentially replicates the current design where a currency is issued by a central bank but private banks hold individuals' savings and process their payments. It is possible to have both systems in parallel, a hybrid CBDC, so the user has a choice.

CBDC advantages: There are several advantages to CBDCs. Firstly, as it is entirely digital it is more efficient than the current solution of paper money, both within and between borders (Bossone & Ardic, 2021). In a one-tier system there is also disintermediation which further increases efficiency (Ahn & Chen, 2022). Secondly, transactions can be tracked which should reduce crime, particularly money laundering and fraud. Thirdly, by offering a CBDC a government reduces the appeal of risky cryptocurrencies and stablecoins. From the user's perspective, opening an account is easier as their government already has the necessary personal details. It should also make having a bank account easier for citizens that live in remote areas as they do not need to travel to a physical bank.

CBDC risks: CBDCs may introduce some risks that must be analysed and understood before an implementation. The three main ones are as follows: (a) Inevitably, a greater reliance on technology makes cybersecurity risks more dangerous. The vulnerabilities can be with the infrastructure providers, the operators of the CBDC and the users (Banxico, 2021). (b) A problematic implementation may lead to low adoption or even reduce trust and confidence in CBDCs. (c) The ease of use of CBDCs may lead to competition between several CBDCs. If a user can easily access several CBDCs they will compare them based on how well they keep their value, as with current currencies. However, unlike current currencies the functionality of the digital wallet may also play a role and become a relative advantage.

## Trust in Currencies, Payments, and Technology

Despite trust typically being based on principles from psychology and sociology that remain relatively consistent, it often has significant differences in a new context where the relationship between the trustor and the trustee is different. For the digital Euro, early signs suggest that image and credibility influence trust (Tronnier et al., 2022). The trustor, in this context a citizen, must trust a government to back the monetary value of the cash they have. In many countries this used to be backed by gold, meaning a citizen in theory could receive gold to the monetary value of the cash they had. This then evolved, leaving gold out of the equation, but the citizen could still hold cash in the form of physical notes if they wanted to. CBDCs once again change this relationship leaving any physical proof out of the equation. This is a fundamental change, and hence 'digital' is in the name of this new form of currency and payment system. While in many countries cash has been on the way out for some time (Tee & Ong, 2016), its role in the relationship with the user, and its role in building trust with the user, should not be underestimated. The new relationship between the trustor (citizen) and trustee (government and central bank) replaces previous physical assurances, with technology. Therefore, trust in the technology is important. Previous research has shown that trust in technology is different to trust in people (Lankton et al., 2015).

In addition to technology, in the relationship between the trustor (citizen) and trustee (government and central bank), there are still people that play a role in the process on the side of the trustee. However, there is a limited direct interaction between the citizen and the people that make up the government and the central bank, and a limited or no personal relationship. Therefore, in this relationship, as with other similar ones, it is better to look at trust in the institutions (Pavlou & Gefen, 2004). The literature therefore suggests that the citizen's trust in a CBDC may be based firstly on trust in institutions and secondly trust in technology. Institutional trust is defined as the consumer's perception that third-party organizations (not the one purchasing or making a sale) can effectively support the exchange (Pavlou & Gefen, 2004).

The term institutional trust has been used since the 1980s to describe the ability of institutions to build trust (Zucker, 1986). For CBDCs institutional trust will depend on the specific risks people and societies face, and which institutions are in an ideal position to build trust back up to the necessary level. These institutions are not just a third party that the consumer is familiar with. There is 'trust transference' between the third party and the trustee. This trust transference is covered by literature on network trust (McEvily et al., 2003), but it is not the only way institutions build trust. The primary way an organization builds institutional trust is by fulfilling a related role to the transaction effectively.

For example, for someone purchasing insurance the institutions that regulate the internet and insurance, are in a position to build trust.

Trust in technology plays a role in the consumer trusting a CBDC because the reduced contact with humans, is replaced with an increased interaction with technology. For example, if a user relies on a CBDC for their payments and they cannot access their electronic wallet due to problems with the Internet connection, or the shop they are trying to pay does not receive the payment, their trust will be reduced. Trust in a technology can be based on different criteria to trust in a person (Lankton et al., 2015). Trust in a system can be influenced by functionality, helpfulness, and reliability (Lankton et al., 2015).

## 6.3 Research Model

Trust has been separated into several different categories by previous research. Two of the most popular distinctions are (1) into people-centric trust and technology-centric trust (Lankton et al., 2015) and (2) institutional trust and technology-focused trust (Pavlou & Gefen, 2004). In a similar way, this research puts forward a model that separates trust in CBDCs into (1) trust in institutions implementing CBDCs and (2) trust in a specific CBDC technology. The four hypotheses related to the influence of trust associated to the institutions involved in CBDCs are presented first, followed by three hypotheses related to the technology used by CBDCs. The eighth and final hypothesis tests whether trust positively influences the intention to use a CBDC.

### Trust Built by Institutions Implementing CBDCs

If institutions directly involved in the implementation of a CBDC such as the government and the central bank are trusted, this will increase trust in the CBDC. The exact role of a government and a central bank may not be the same in all cases, but in the cases implemented so far and the main ones proposed, both have a role. For example, an EU and UK implementation involves different responsibilities for the respective governments and central banks, but in both cases both institutions are involved (Morgan, 2022; Mooij, 2022). Typically, a government will pass certain related laws, and the central bank will implement and run the operations of a CBDC. Some countries' central banks, or equivalent, have more independence from the government, but in most cases, there is close cooperation. For a historical decision such as implementing a CBDC there is no evidence to this day of governments not working closely with their central banks. Therefore, the first hypothesis is:

H1: Trust in the government and the central bank issuing a CBDC will increase trust in the CBDC.

A guarantee is a formal assurance in writing that something will happen. Guarantees have a history of being used to reassure users of technology or consumers of financial services, and they are often beneficial (Martínez-López et al., 2020). Specific guarantees offered to the user of a CBDC will increase trust in the CBDC. The guarantees can be on any aspect of the CBDC including the currency retaining its value or it being able to complete all the forms of payment necessary. Therefore, the second hypothesis is:

H2: Expressed guarantees for the user of a CBDC will increase trust in the CBDC.

Personal information privacy concern is caused when consumers must share sensitive personal information to receive a product or service (Yun et al., 2019). New privacy concerns can emerge each additional time this personal information must be shared with a new organization (Dinev et al., 2013). Evidence of this is the extent to which consumers will provide false personal information to avoid revealing their genuine personal information (Miltgen & Jeff Smith, 2019). As the government already holds personal information of the consumer, this might increase trust. Therefore, the third hypothesis is:

H3: Personal data handled when operating a CBDC by a government, that already holds personal information of the user, will increase trust in the CBDC.

While the trust in the organizations implementing the technology is clearly important, it is not sufficient on its own. This technology, as most technologies, has been implemented for some time before the consumer is considering adopting it. During this time a reputation has been built. Reputation has been proven to influence trust in a variety of contexts involving technology (Einwiller, 2003), including trust in financial services (Dupont & Karpoff, 2020). Therefore, it is hypothesized that:

H4: The positive reputation of existing CBDCs implemented in other countries will increase trust in the CBDC.

## Trust Built by the Specific CBDC Technology Implemented

The process of using a CBDC is digital and over the Internet, so the technology, including blockchain, handles the necessary processes (Bossone & Ardic, 2021). This automation reduces human involvement and corruption (Ahn & Chen, 2022). As human involvement is reduced, so is the need to trust humans. Therefore, it is hypothesized that:

H5: The automation and reduced human involvement achieved by a CBDC technology increases trust.

A service can include trust-building functionality such as third-party certification, and policies that protect the consumer (Chang et al., 2013). For example, in the European Union like many other parts of the world, there are policies protecting the consumer's bank savings, up to a certain point (Chiaramonte et al., 2020). Similarly, the technology of a CBDC and the policies around it can have specific trust-building features. These can include a well-designed interphase and two-factor authentication to give the user a strong sense that they are in control of their money (Reese et al., 2019). Therefore, it is hypothesized that:

H6: The trust-building functionality of a CBDC wallet app will increase trust in the CBDC.

Despite the large volume of data provided by a typical consumer so that they can receive the products and services they want, this does not happen without some privacy concerns (Gu et al., 2022). CBDCs have already generated some privacy concerns (Pocher & Veneris, 2022) despite not being widely available yet for most people. Additional privacy features of the CBDC wallet app and back-end processes, such as anonymity, can reduce personal information privacy concern (Dinev et al., 2013). Therefore, it is hypothesized that:

H7: Privacy features of the CBDC wallet app and back-end processes such as anonymity will increase trust in the CBDC.

The previous seven hypotheses identified the institutional and technological ways trust is built. The eighth and final hypothesis attempts to verify that consumer trust in a CBDC does, indeed, encourage the use of CBDCs. The originality and contribution of this research lies primarily in the previous seven hypotheses, but the final hypothesis is necessary to confirm that trust plays a role in this context as it does in many other similar contexts (Lankton et al., 2015; McKnight & Chervany, 2002). Therefore, the final hypothesis is:

H8: Trust in a CBDC will increase the willingness to use a CBDC.

The initial research model on how trust in a CBDC is built in seven ways, and how trust in CBDCs increased the willingness to use them, is illustrated in Fig. 6.1.

## 6.4   Method

### Data Collection

The data for the quantitative analysis was collected by online survey hosted on the SoSci Survey tool (www.soscisurvey.de). The survey items are based on measures validated in similar contexts and use a seven-point Likert-type scale. The classic Likert-type scale started from 1, strongly disagree, up to 7, strongly agree. Each latent variable was measured by three measured variables, as illustrated in Table 6.1. The survey was designed to take less than ten minutes to complete. These ten minutes included the time needed to read the instructions, complete the demographic questions and those related to the model.

The minimum sample size necessary was calculated based on the guidelines for SmartPLS (Hair et al., 2021). Based on the maximum number of arrows pointing to a latent variable being seven, the minimum sample size is 228, for a statistical power of 80% (Hair et al., 2021). The minimum number was exceeded as after 429 were collected, 388 were considered to be valid. Using a higher number than the minimum increases the precision of the estimations (Hair et al., 2021). Of the 429 surveys collected, 41 raised a red flag and did not pass the survey quality checks. Firstly, there was a question at the start asking the participant to select a definition of what a CBDC is, to test if they had sufficient knowledge to give useful answers. Additionally, there was a question that was asked twice with a different wording to check if conflicting answers were given. Surveys were also rejected for other typical reasons such as being incomplete, completed too fast known as 'speeder flag', and for the same response being selected for all the questions known as 'straight-lining flag'. The demographic analysis of the 388 valid surveys completed, presented in Table 6.2, show a good balance between genders, age groups, education levels, and incomes.

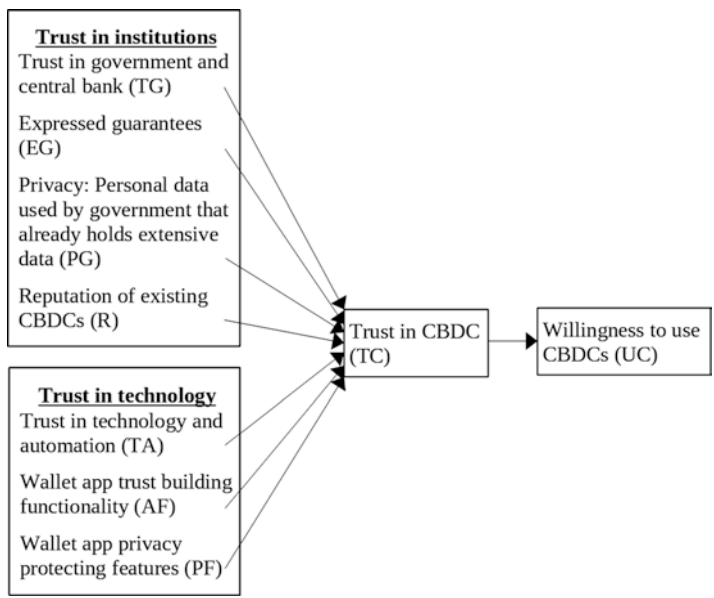

**Fig. 6.1** Initial research model on how trust in a CBDC is built in seven ways

**Table 6.1** Latent variables and their measures

| Latent variables | Measures | Source of construct measures adapted |
|---|---|---|
| Trust in government and central bank (TG) | TG1, TG2, TG3 | Grimmelikhuijsen and Knies (2017) |
| Expressed guarantees (EG) | EG1, EG2, EG3 | Martínez-López et al. (2020) |
| Privacy: Personal data used by government that already holds extensive data (PG) | PG1, PG2, PG3 | Dinev et al. (2013), and Yun et al. (2019) |
| Reputation of existing CBDCs (R) | R1, R2, R3 | Einwiller (2003) |
| Trust in technology and automation (TA) | TA1, TA2, TA3 | Lankton et al. (2015), McKnight and Chervany (2002) |
| Wallet app trust-building functionality (AF) | AF1, AF2, AF3 | Pavlou (2002) |
| Wallet app privacy-protecting features (PF) | PF1, PF2, PF3 | Dinev et al. (2013) |
| Trust in CBDC (TC) | TC1, TC2, TC3 | Lankton et al. (2015), McKnight and Chervany (2002) |
| Intention to use CBDC (UC) | UC1, UC2, UC3 | Venkatesh et al. (2003) |

## Data Analysis Technique

The quantitative method evaluated the model developed using Structural Equation Modelling (SEM) with the variance-based Partial Least Squares (PLS) approach. This is an Ordinary Least Squares regression method. An initial model was tested and improved so that a model with strong empirical and theoretical support is generated. The SmartPLS 4.0 software was used to explore and evaluate the model. The PLS algorithm was set to run for 3000 iterations. Because the model has a degree of complexity with three tiers of latent variables, the analysis first evaluated the measurement model, followed by the evaluation of the structural model. The reflective measurement model evaluates the relationship between the measured variables and their respective latent variables, while the structural model evaluates the relationship between the latent variables themselves (Hair et al., 2021). It is necessary to verify that the measured variables do indeed represent the latent variables, before moving on to exploring the relationship between the latent variables.

**Table 6.2** Demographic information of the survey sample group

| Measure | Category | Frequency |
|---|---|---|
| Gender | Female | 222 |
| | Male | 166 |
| Age | Younger than 18 | 0 |
| | 18–24 | 147 |
| | 25–44 | 126 |
| | 45–64 | 84 |
| | 64 or older | 31 |
| Education level | No high school education | 4 |
| | High school graduate | 193 |
| | University bachelor degree | 154 |
| | University master or doctoral degree | 37 |
| Income | No regular income | 48 |
| (in Euro per month) | 400–1200 | 57 |
| | 1201–3000 | 147 |
| | 3001–5000 | 110 |
| | More than 5000 | 26 |

## 6.5   Analysis and Results

### Measurement Model

Before the relationships between the latent variables can be tested, the relationship between each latent variable and its respective measured variables must be tested to see if it meets the necessary criteria (Hair et al., 2021). The first thing to be evaluated are the measured variables' factor loadings. These are above the minimum required level of 0.7 with the lowest being 0.864. The Average Variance Expected (AVE) should be above 0.5, and this requirement is met by all of them as the lowest is 0.807. Therefore, both the factor loadings and AVE meet the requirements and show a sufficient level of convergent validity. The Composite Reliability (CR) is above 0.7 for all the variables with the lowest value being 0.883. This indicates that there is adequate internal consistency and individual construct reliability, between the latent variable and the measured variables. The discriminant validity, presented in Table 6.3 evaluated using the Fornell-Larcker criterion, shows that the measured variables have a stronger relationship with their own latent variable compared to any of the other variables. Based on the analysis discussed above and present in Table 6.3, the measurement model is supported, and the analysis can move on to the structural model.

**Table 6.3** Results of the measurement model analysis

| Variables | | Loadings | CR | AVE | Discriminant validity (Fornell-Larcker criterion) | | | | | | | | |
|---|---|---|---|---|---|---|---|---|---|---|---|---|---|
| | | | | | TG | EG | PG | R | TA | AF | PF | TC | UC |
| TG | TG1 | 0.949 | 0.948 | 0.905 | 0.951 | | | | | | | | |
| | TG2 | 0.953 | | | | | | | | | | | |
| | TG3 | 0.952 | | | | | | | | | | | |
| EG | EG1 | 0.899 | 0.913 | 0.851 | 0.893 | 0.923 | | | | | | | |
| | EG2 | 0.954 | | | | | | | | | | | |
| | EG3 | 0.914 | | | | | | | | | | | |
| PG | PG1 | 0.920 | 0.887 | 0.814 | 0.900 | 0.848 | 0.902 | | | | | | |
| | PG2 | 0.896 | | | | | | | | | | | |
| | PG3 | 0.891 | | | | | | | | | | | |
| R | R1 | 0.926 | 0.883 | 0.807 | 0.812 | 0.862 | 0.794 | 0.898 | | | | | |
| | R2 | 0.904 | | | | | | | | | | | |
| | R3 | 0.864 | | | | | | | | | | | |
| TA | TA1 | 0.934 | 0.916 | 0.853 | 0.875 | 0.834 | 0.872 | 0.808 | 0.924 | | | | |
| | TA2 | 0.918 | | | | | | | | | | | |
| | TA3 | 0.919 | | | | | | | | | | | |
| AF | AF1 | 0.893 | 0.889 | 0.818 | 0.713 | 0.754 | 0.708 | 0.748 | 0.679 | 0.905 | | | |
| | AF2 | 0.898 | | | | | | | | | | | |
| | AF3 | 0.922 | | | | | | | | | | | |
| PF | PF1 | 0.928 | 0.899 | 0.828 | 0.647 | 0.703 | 0.642 | 0.733 | 0.640 | 0.883 | 0.910 | | |
| | PF2 | 0.921 | | | | | | | | | | | |
| | PF3 | 0.880 | | | | | | | | | | | |
| TC | TC1 | 0.916 | 0.867 | 0.789 | 0.857 | 0.875 | 0.835 | 0.857 | 0.855 | 0.844 | 0.823 | 0.904 | |
| | TC2 | 0.869 | | | | | | | | | | | |
| | TC3 | 0.880 | | | | | | | | | | | |
| UC | UC1 | 0.897 | 0.882 | 0.807 | 0.803 | 0.821 | 0.770 | 0.822 | 0.792 | 0.796 | 0.783 | 0.888 | 0.899 |
| | UC2 | 0.915 | | | | | | | | | | | |
| | UC3 | 0.884 | | | | | | | | | | | |

## Structural Model

The structural model, also referred to as the inner model, evaluates the relationship between the latent variables, also known as constructs (Hair et al., 2021). As with the measurement model, several criteria must be met. The coefficient of determination R2 for the endogenous latent variable TC is 0.898, and for the second endogenous variable UC is 0.817. Both are above 0.67 and can be considered 'substantial' (Chin, 1998). The effect size ($f2$) for the paths TG-TC (0.020), EG-TC (0.035), R-TC (0.025), TA-TC (0.107), AF-TC (0.048), PF-TC (0.111) are weak but significant. TC-UC (4.460) is strong. The effect size of the path PG-TC (0.000) can be considered insignificant. Effect sizes are typically interpreted in the following way: insignificant under 0.02, weak between 0.02 and 0.15, moderate between 0.15 to 0.35 and strong above 0.35 (Chin, 1998). PLS-SEM is not focused on model fit, and there is a debate as to whether the estimations of fit it produces should be reported (Hair et al., 2021). However, as an additional indication, with the reservations noted, the Standardized Root-Mean Residual (SRMR) for a saturated model is 0.051. Values below 0.08 indicate a good model fit (Hu & Bentler, 1999). The structural model was further explored with the bootstrapping method set to 5000 samples, and the results were similar, as illustrated in Table 6.4. We see that there is support for all the relationships apart from PG-TC.

There are three typical approaches to SEM analysis. The first is to test a model and the second is to compare alternative models. The third approach, which is followed here, is to generate a model. This is achieved by first proposing an initial model and then making adjustments to improve the statistical support for it. Given the insufficient support for the relationship PG-TC, this variable was removed, and the PLS algorithm and Bootstrapping tests were repeated. The final supported model with one variable omitted is still logical and supported by the literature.

**Table 6.4** Results of the initial structural model

| Path | Sample mean | Standard deviation | T statistics | *p*-value |
|------|-------------|--------------------|--------------|----------|
| TG-TC | 0.127 | 0.052 | 2.436 | 0.015 |
| EG-TC | 0.160 | 0.044 | 3.606 | 0.001 |
| PG-TC | 0.012 | 0.051 | 0.318 | 0.750 |
| R-TC | 0.113 | 0.039 | 2.869 | 0.004 |
| TA-TC | 0.246 | 0.044 | 5.510 | 0.001 |
| AF-TC | 0.164 | 0.047 | 3.546 | 0.001 |
| PF-TC | 0.236 | 0.040 | 5.850 | 0.001 |
| TC-UC | 0.904 | 0.012 | 72.768 | 0.001 |

The effect size ($f2$) for the paths TG-TC (0.026), EG-TC (0.036), R-TC (0.026), TA-TC (0.126), AF-TC (0.050), PF-TC (0.111) are weak but significant. TC-UC remains strong as it is not affected by removing the variable PG. Therefore, removing the variable PG has increased the effect sizes of the remaining variables and thus improved the model. The model fit did not change significantly as the SRMR for a saturated model remains at 0.051. The SRMR for the estimated model had an insignificant change from 0.053 to 0.052. As mentioned earlier, values below 0.08 indicate a good model fit (Hu & Bentler, 1999). The structural model was further explored with the bootstrapping method set to 5000 samples, and the results were similar as illustrated in Table 6.5.

## 6.6    Findings

Several countries have plans to implement CBDCs, while many others are exploring this option (Xu, 2022; Banxico, 2021). The starting point of this research was reviewing the literature and identifying both the potential benefits of CBDCs and the importance of consumer trust in its adoption. An initial model recommended seven ways trust in a CBDC is built. This was evaluated and further developed using Structural Equation Modelling (SEM). The fourth hypothesis, which is the only one not supported, states: 'Personal data handled when operating a CBDC by a government, that already holds personal information of the user, will increase trust in the CBDC'. It appears that users of a CBDC do not see it as an advantage that their personal information is shared with an organization that already has extensive information on them. The final model, with six ways trust in a CBDC is built, is supported by the data and literature and is illustrated in Fig. 6.2.

**Table 6.5** Results of the final structural model

| Path | Sample mean | Standard deviation | T statistics | *p*-value |
|------|-------------|--------------------|--------------|-----------|
| TG-TC | 0.127 | 0.052 | 2.436 | 0.015 |
| EG-TC | 0.160 | 0.044 | 3.606 | 0.001 |
| R-TC | 0.113 | 0.039 | 2.869 | 0.004 |
| TA-TC | 0.246 | 0.044 | 5.510 | 0.001 |
| AF-TC | 0.164 | 0.047 | 3.546 | 0.001 |
| PF-TC | 0.236 | 0.040 | 5.850 | 0.001 |
| TC-UC | 0.904 | 0.012 | 72.768 | 0.001 |

## Theoretical Contribution

This research makes three primary theoretical contributions: (1) The first contribution is the model of trust in CBDCs that identifies six ways to build trust in a CBDC. (2) The second contribution is that the model supports the importance of consumer trust in the adoption of a CBDC. (3) Thirdly, this research links the literature on trust to CBDCs and highlights the benefits of exiting literature to understanding CBDCs and overcoming their challenges to adoption.

From the six approaches to building trust, the first three apply to trust in the institutions involved, while the final three apply to trust in the technology used. The first three that apply to trust in the institutions involved are as follows: (1) Trust in government and central bank, (2) expressed guarantees for the user, and (3) the positive reputation of existing CBDCs active elsewhere. The final three that apply to trust in the technology used are thus: (4) The automation and reduced human involvement achieved by a CBDC technology, (5) the trust-building functionality of a CBDC wallet app, and (6) privacy features of the CBDC wallet app and back-end processes such as anonymity.

These six methods to build trust in CBDCs now extend the literature they were based on to CBDCs. The literature on institutional trust (Pavlou & Gefen, 2004), the benefits of guarantees to building trust (Martínez-López et al., 2020), and the role of reputation in building trust (Dupont & Karpoff, 2020) are extended to CBDCs. Similarly, the literature of the three more technology-focused variables is extended to CBDCs. These are automation and AI (Ahn & Chen, 2022), trust-building functionality in technology (Chang et al., 2013), and privacy features in a technology (Dinev et al., 2013).

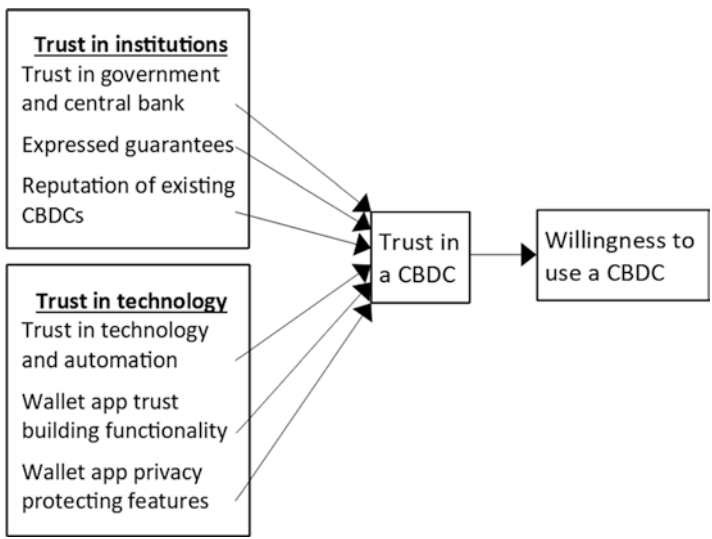

**Fig. 6.2** Final, supported research model on how trust in a CBDC is built in six ways

## Practical Contribution

This research has practical implications for the various stakeholders involved in implementing and operating a CBDC but also the stakeholders in the ecosystem using a CBDC. The stakeholders involved in delivering and operating CBDCs such as governments, central banks, regulators, retail banks, and technology providers can apply the six trust-building approaches so that the consumer trusts a CBDC sufficiently and adopts them. The many organizations that are not directly involved in operating the CBDC, but are affected by it, such as retailers can also build trust with some of the six methods. The six methods may not apply to all the stakeholders, and a stakeholder may not be able to implement all six, but they can identify those that do apply.

## 6.7   Conclusion, Limitations, and Future Research

CBDCs are an important part of the new Fintech solutions disrupting finance, but also more generally society. This research evaluated a model on how trust in a CBDC is built. Data was collected by survey and analysed with PLS-SEM. This research verified the importance of trust in CBDC adoption and identified six ways to build trust in CBDCs. These are (1) trust in government and central bank issuing the CBDC, (2) expressed guarantees for the user, (3) the positive reputation of existing CBDCs active elsewhere, (4) the automation and reduced human involvement achieved by a CBDC technology, (5) the trust-building functionality of a CBDC wallet app, and (6) privacy features of the CBDC wallet app and back-end processes such as anonymity. The first three trust-building methods relate to trust in the institutions involved, while the final three relate to trust in the technology used. Trust in the technology is like the walls of a new building, and institutional trust is like the buttresses that support it.

This research has practical implications for the various stakeholders involved in implementing and operating a CBDC but also the stakeholders in the ecosystem using CBDCs. The stakeholders involved in delivering and operating CBDCs such as governments, central banks, regulators, retail banks, and technology providers can apply the six trust-building approaches so that the consumer trusts a CBDC and adopts it.

The limitations of this research open new paths for future research. The use of CBDCs is still at a nascent stage. Firstly, while the few existing and many planned implementations give us a sufficiently stable foundation to build on, the findings should be taken in the context of a technology in the early stages of adoption. Secondly, the sample was from one country, Germany, so the findings can be tested in other countries and regions of the world. Thirdly, in the future, new measures can be taken to build trust in CBDCs, so this space should be reviewed regularly. Finally, this research recommends that Fintech is not only treated as an interdisciplinary topic, as it rightly is until now, but also as a distinct field of research that needs specialized research that captures its idiosyncrasies. The full impact of Fintech may not be captured just by repurposing theory from adjacent fields like finance and information systems.